\documentclass[a4paper,11pt]{article}
\usepackage{amsmath}
\usepackage{amssymb}
\usepackage{amsthm}
\usepackage{enumitem}
\usepackage{jinstpub} 
\usepackage{multirow}

\pdfoutput=1          

\title{\boldmath Heat radiation reduction in the cryostat with multilayer insulation technique}

\author[a,1]{D. Singh,\note{Corresponding author.}}
\author[a,]{A. Pandey,}
\author[a,c]{M. K. Singh,}
\author[b]{L. Singh,}
\author[a,b,2]{V. Singh,\note{Corresponding author.}}

\affiliation[a]{Banaras Hindu University, Department of Physics, Institute of Science\\Varanasi$-$221005, India}
\affiliation[b]{Central University of South Bihar, Physics Department, School of Physical and Chemical Sciences\\Gaya$-$824236, India}
\affiliation[c]{Institute of Physics, Academia Sinica\\Taipei$-$11529, Taiwan}
\emailAdd{daminisinghphy27@gmail.com}
\emailAdd{venkaz@yahoo.com}
\abstract{Multilayer insulation (MLI) is an important technique for the reduction of radiation 
  heat load in cryostats. The present work is focused on investigation for the selection of 
  suitable reflective layer and spacer material in MLI systems. In our analysis, we have 
  selected perforated double$-$Aluminized Mylar (DAM) with Dacron, unperforated DAM with Silk$-$net 
  and perforated DAM with Glass$-$tissue for their evaluation as the reflective layer as well as 
  spacer materials in MLI technique. Current work would discuss the calculation of the 
  effect of layer density and the number of layers on the heat load. Knowing the key parameters 
  of MLI, we have compared the heat load generation in spherical as well as cylindrical 
  cryostats and the effect of layering near and outer surface on the heat load.}

\begin{document}
\maketitle
\flushbottom

\section{Introduction}
\label{sec:intro}

\begin{figure}[htbp]
\centering 
\includegraphics[width=6.6cm]{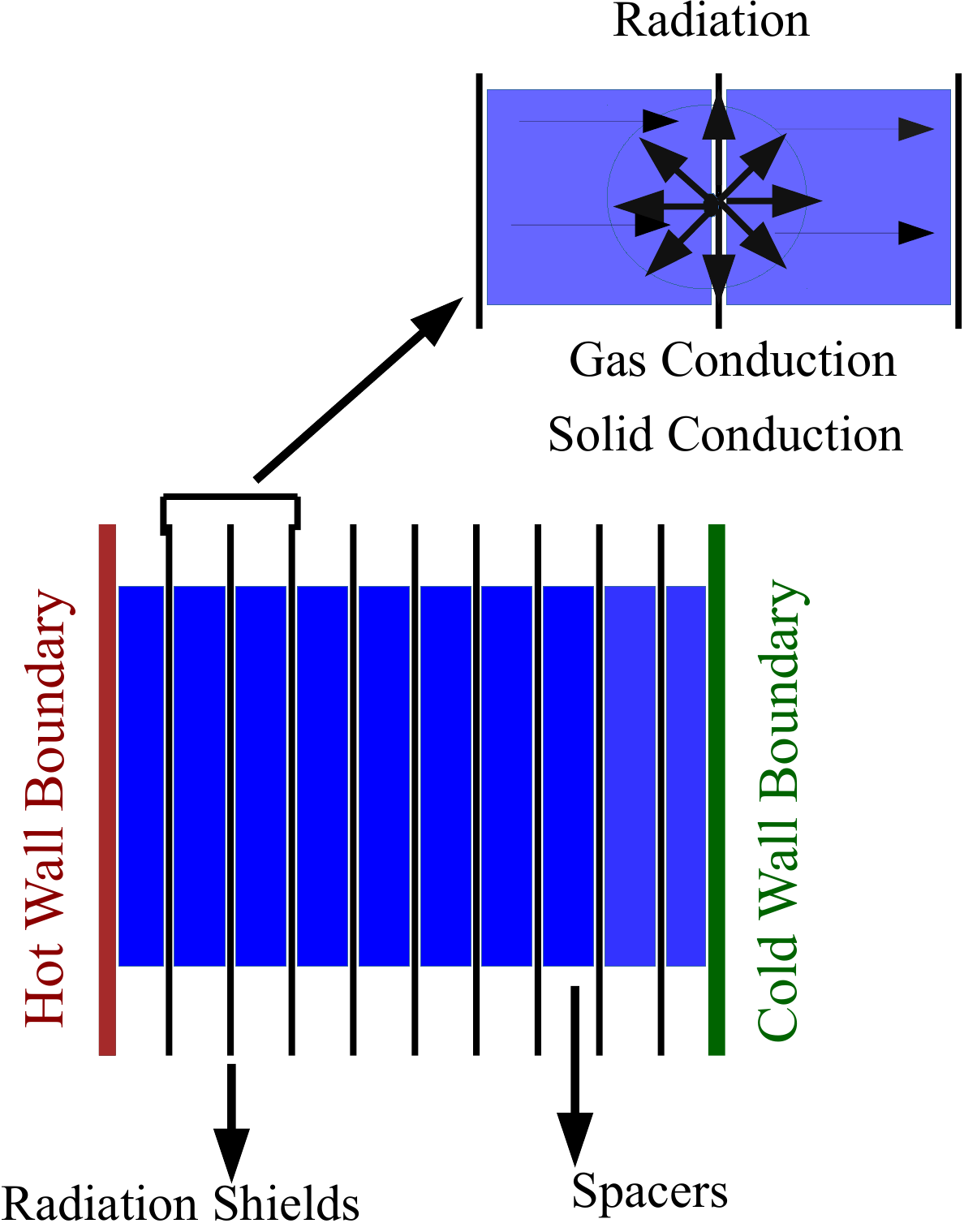} 
\caption{\label{fig:MLI} Schematic diagram of the MLI structure consists of 
reflective layers interleaved with insulating spacers. The heat exchange (via 
heat radiation, solid conduction and gas conduction) between two consecutive 
reflective layers is depicted in its inset.}
\end{figure}

The word cryostat (also called as ``dewars'' in the memories of Sir James Dewar) 
is made up of two words cryo and stat, which means cold and stable~\cite{YBC:Book:2016}. 
Typically, it is a container filled with cryogenic liquid (Liquid Argon (LAr), 
Liquid Helium (LHe), Liquid Nitrogen (LN$_{2}$), and Liquid Hydrogen (LH$_{2}$) etc.) to provide 
mechanical housing, cooling and shielding against the residual environmental backgrounds 
to the device under consideration at very low temperature for its safe and 
stable operation~\cite{V_Parma}. Evidently, steady functioning requires minimum heat 
load (thermal radiation, solid conduction and gas conduction) near the inner wall of 
the cryostat. The requirement of minimum heat load (Particularly radiation heat load) 
can be well accomplished by using the multilayer insulation (MLI) technique. The basic 
principle of MLI technique is to obtain the multiple radiation 
reflection by placing the reflective layers called as radiation shields, in the vacuum 
space between the two walls (hot radiating surface and cold surface) of the cryostat~\cite{JME:Book:2012}. 
These reflective layers are usually made up of thin polyethylene or Mylar sheet, 
coated with highly reflecting material (Aluminium or Gold, but most commonly Aluminium 
is used due to low cost) on both sides~\cite{hedayat:2002}. It follows that there may 
be a chance of conduction due to adjacent reflective layers. Therefore, low conductivity 
materials or insulators called as spacers are placed in between these reflective layers. 
As these reflective layers are interleaved with insulating spacers, they do not 
touch each other and minimize the thermal heat exchange/thermal conduction~\cite{Sutheesh:2018}. 
A schematic diagram of the MLI structure consists of reflective layers interleaved with 
insulating spacers is shown in figure~\ref{fig:MLI}. \par 

Vigorous thermal insulation systems of MLI technique are required for 
developing the efficient storage and transfer of cryogens~\cite{NASA:2008}. It is 
a passive thermal protection system widely used in cryogenics and space exploration 
programs as an excellent thermal insulator~\cite{Sutheesh:2018}. MLI technique has extensive 
applications in storage, transfer, thermal protection, and low$-$temperature processes. 
The present work focuses on thermal protection and low$-$temperature applications of 
MLI technique in fundamental physics research experiments. MLI technique is being used worldwide in 
numerous basic physics research experiments such as: exploration of space 
(NASA~\cite{NASA:2008, JEF:2018}), accelerators (LHC~\cite{V_Parma, Baudouy:2015}), 
dark matter searches (EDELWEISS~\cite{EDELWEISS:2011, EDELWEISS:2012}, 
CRESST~\cite{CRESST:20121, CRESST:20122}, EURECA~\cite{EURECA:2014}), and the searches 
of neutrinoless double $\beta$-decay (CUORE~\cite{CUORE:2018}, GERDA~\cite{GERDA:2005, GERDA:2006}, 
and in future LEGEND~\cite{LEGEND}), etc. These broad fields of 
applications reveal the enthusiasm of the scientists working in this field 
worldwide. There are ample cryogenic thermal research works being performed after 
the first experimental test on MLI by Sir James Dewar in 1900 when he tested with
three layers of Aluminum foil (current form of MLI with layered radiation shield 
system was described firstly by William D. Cornell in the late 
1940s)~\cite{Johnson:2010, Cornell:1947}. \par

The design, analysis, and testing of a given MLI system depend on the nature and 
requirements of the specific application. Generally, MLI systems are used in high 
vacuum environments, which may be a vacuum vessel or space such as the Earth's 
orbit or the lunar surface~\cite{JEF:2018}. This is because MLI is the ideal 
insulation for the radiation dominated heat transfer environments such as spacecraft 
in the low Earth orbit and the other high vacuum functions~\cite{WLJ:2014}. 
Furthermore, MLI systems can also be designed for soft vacuum purposes in industrial 
products as well as in commercial building apparatus~\cite{SDA:Book:2000}. 
Nevertheless, MLI systems can also be constructed for no vacuum (at ambient pressure) 
applications where radiation heat transfer is a significant amount of the total 
heat gain in moderate temperature (non$-$cryogenic) thermal protection 
systems~\cite{JEF:2018}. Tremendously low temperature ($\leq$~4~K) refrigeration is 
also attainable by MLI systems, which have numerous worldwide applications in 
basic physics research~\cite{Sutheesh:2018}. \par

The current work is an attempt to address the question of what is the best MLI 
system, which is usually asked before designing the cryostat in an experiment. Although
reducing the heat load in a system is the primary aim, it should not be too much costly
as well as should not take too much time in operations. It follows that the correct 
methodology would be to proceed via an appropriate ``thermo$-$economic'' approach in achieving the 
best MLI system. In this direction, the selection of suitable material and appropriate 
design are very crucial. Furthermore, the level of vacuum (incorporating the level of 
vacuum between the layers also) and the layer density are even more pivotal for getting
the better efficiency of the MLI system. The present work is focused on the selection of 
suitable material with optimum layer density to minimize the heat/thermal radiation 
(the most prominent contribution in the total thermal budget for cryostats) and achieve 
the best MLI system for the spherical and cylindrical cryostats. \par

\section{Heat exchange between the two surfaces of a cryostat}
\label{sec:heat_transfer}
Cryostats are made up of metals and due to environmental effects, there are always some 
possibilities for the production and exchange of heat between its surfaces. There are 
three types of heat load at very low temperature that might take place in a cryostat: 
(1) \textit{Solid conduction} is a type of heat transfer in which there exists a 
temperature gradient between the two solid surfaces in the supporting systems. (2) 
\textit{Residual gas conduction} is a type of heat transfer that occurs between a 
surface and a moving liquid when they are at different temperatures of non$-$perfect 
vacuum insulation and (3) \textit{Thermal radiation} is a mode of the heat transfer 
between two surfaces at different temperatures in the absence of media, which means 
it can propagate in vacuum too~\cite{V_Parma}. In fact, thermal radiation is the most important 
and major part of the total produced heat load in the cryostats. All of these heat 
loads can be minimized in various ways by following appropriate procedures. Solid 
conduction can be minimized by making a suitable and appropriate choice of the 
material. Residual gas conduction can be minimized by creating a perfect vacuum in 
between the walls of the cryostat. Thermal radiation can be minimized by placing the 
radiation shields with spacers in between the cold and hot surfaces of a cryostat. 
Therefore, understanding of the thermal radiation reduction requires appropriate 
consideration of the heat transfer between these two surfaces~\cite{JGW:Book:2016}. 
The present work focuses on the reduction of thermal radiation heat load. \par

In the minimization of thermal radiation, an opaque body would be a good approximation 
in designing the cryostat structure. This is because an opaque body has zero$-$transmissivity 
to the electromagnetic radiation incident over it. It follows that all 
the received energy is either absorbed or reflected from the surface of such a body. 
However, this approximation fails in the presence of an orifice in the thermal shield. 
Such hole or crack leads shining of radiation through them and finally, these radiations 
get absorbed by the internal surfaces via multiple reflections. Then the body will 
behave like a black body with full absorptivity. Therefore, such gaps, holes and, 
slots must be avoided by taking great care during the designing of thermal shields, 
which may be pernicious to the thermal performance of cryostats~\cite{V_Parma}. If 
such gaps are unavoidable (when thermal contraction compensation gaps need to be surely 
introduced), then MLI blankets (reflective layer + spacer) can be used to avoid shining 
light through them. However, in ultra$-$high vacuum applications (where MLI blankets 
can't be used) such gaps need to be surrounded by the traps of high$-$absorptivity and 
high$-$reflectivity materials (by making special coatings) to absorb and reflect the 
light incident on the wall of thermal shield and reduce the multi$-$path reflection on 
the inside surface. \par

This problem can be well understood by considering the radiation exchange between 
two surfaces. The heat exchange between the real surfaces of two bodies (with area 
$A_{1}$ and $A_{2}$ which are kept at temperatures $T_{1}$ and $T_{2}$ ($T_{1} > T_{2}$), respectively) 
greatly depends on their emissivity ($\varepsilon$) which varies with the temperature 
and defines as the fraction of the emitted radiation $E(T)$ with respect to that of a 
black body $E_{b}(T)$~\cite{V_Parma}
\begin{equation}
\label{EQ:heatExch}
q_{1-2} = \varepsilon\sigma(T_{1}^{4} - T_{2}^{4})A_{2}F_{21}~~\rm{where}~~\varepsilon = \left[\frac{E(T)}{E_{b}(T)}\right] \leq 1.
\end{equation}
Here $\sigma$ is the Stefan$-$Boltzmann's constant (5.675$\times10^{-8}$ Wm$^{-2}$K$^{-4}$) 
and $F_{21}$ is the geometrical view factor~\cite{MFMd:2013, WEB}, defined as the fraction 
of the total radiation leaving from the first body, which is intercepted and absorbed by 
the second body (it depends upon the relative orientation of the two surfaces). It 
follows the reciprocity rule such that $A_{1}F_{12} = A_{2}F_{21}$~\cite{FPIn:2002}. In case 
of metals at cryogenic temperatures, the $\varepsilon$ reduces $\sim\propto T$. This 
leads to the enhancement in the low$-$emissivity properties of cryogenic cooled thermal 
shields in cryostats. \par
\begin{figure}[htbp]
\centering 
\includegraphics[width=12.0cm]{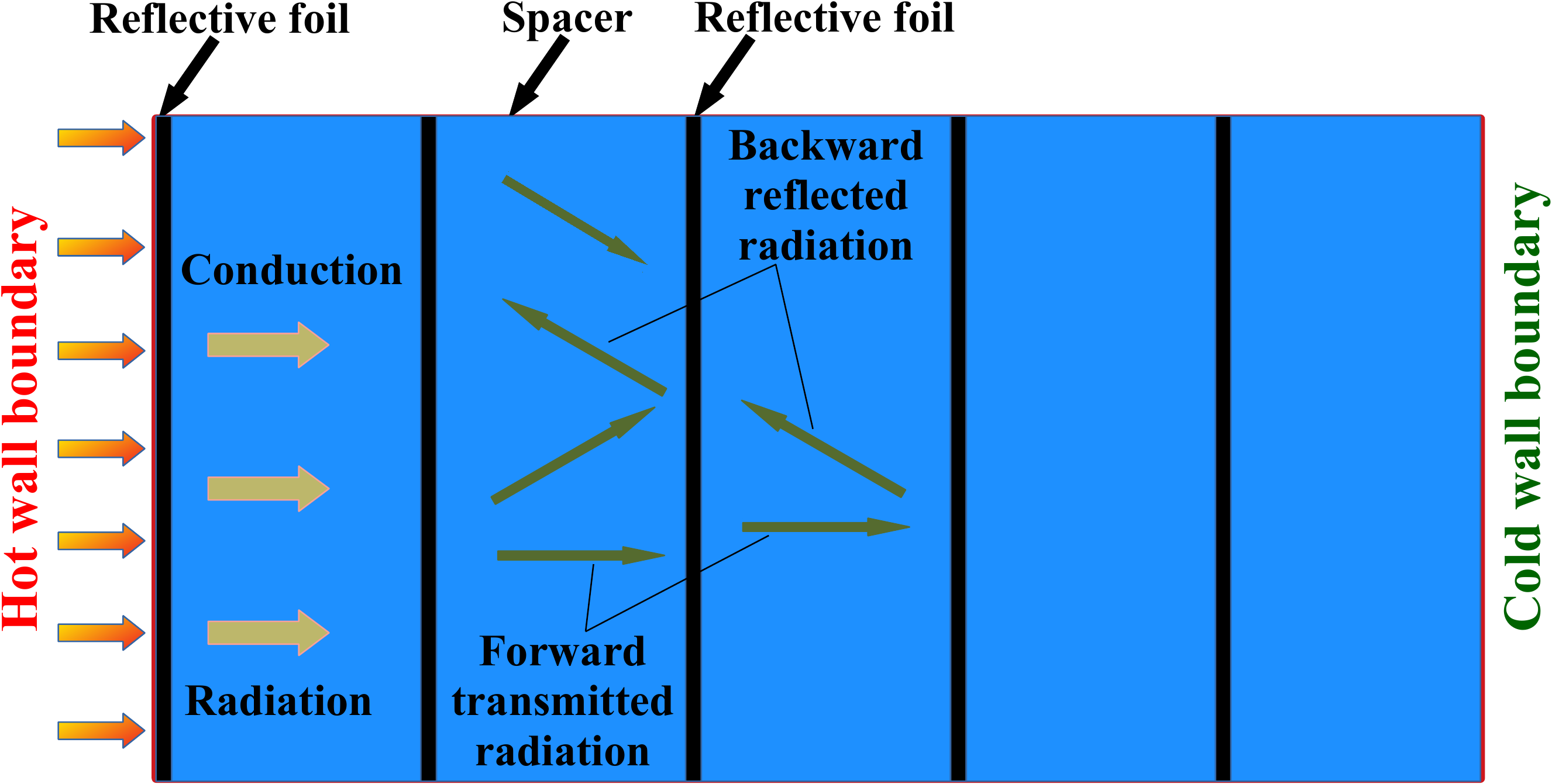} 
\caption{\label{fig:MLI_Function} Basic functioning of MLI technique. Heat load tremendously 
  decreases as the heat radiation move on to the consecutive reflective layers.}
\end{figure}
\begin{figure}[htbp]
\centering 
\includegraphics[width=11.0cm]{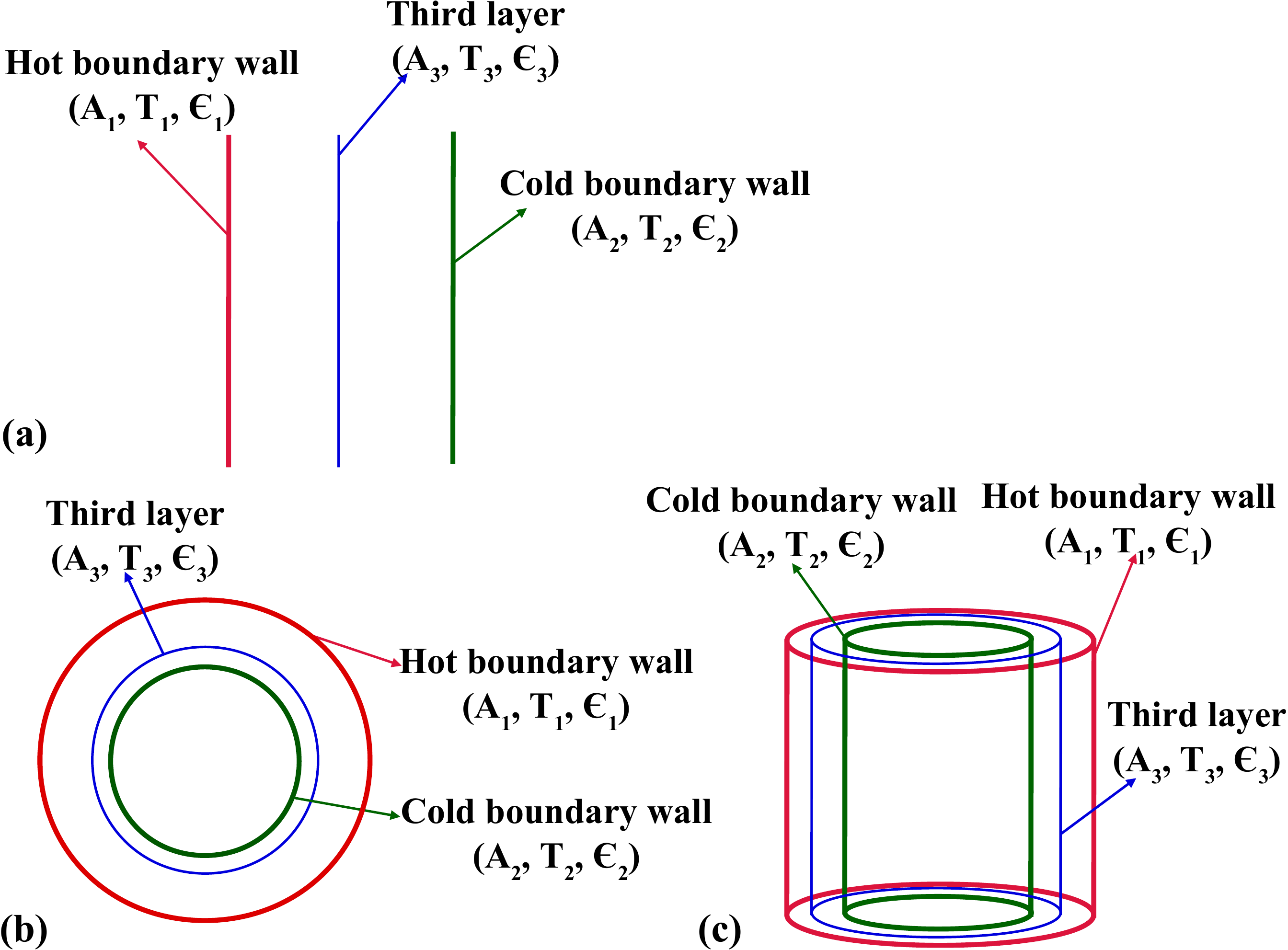} 
\caption{\label{fig:third_layer} Schematic diagrams of (a) parallel plate (b) spherical, and 
  (c) cylindrical geometries of a cryostat with the insertion of a third intermediate layer.}
\end{figure}

It is the geometries which play a crucial role in the designing of cryostats. Generally, the heat 
transfer by radiation between two enclosed surfaces (from one which is at $T_{1}$ with 
$\varepsilon_{1}$ and surface area $A_{1}$, to another at $T_{2}$ with $\varepsilon_{2}$ 
and $A_{2}$, such that $T_{1} > T_{2}$) can be expressed as~\cite{Baudouy:2015}
\begin{equation}
\label{EQ:General}
q_{1-2} = \frac{\sigma(T_{1}^{4} - T_{2}^{4})}{\left(\frac{1-\varepsilon_{1}}{\varepsilon_{1}A_{1}} + \frac{1}{A_{2}F_{21}} + \frac{1-\varepsilon_{2}}{\varepsilon_{2}A_{2}}\right)}~~.    
\end{equation}
The present work focuses on the comparative study of radiation heat load calculation in 
the cylindrical and spherical designs. The heat exchange between two enclosed cylinders with 
$A_{1}$ and $A_{2}$ being the outer and inner surfaces respectively, and kept at temperatures 
$T_{1} > T_{2}$, $F_{21} = 1$ (because surface $A_{2}$ is completely surrounded by the surface 
$A_{1}$~\cite{WEB:Geometric} as shown in figure~\ref{fig:third_layer}(c)), from energy balance 
and use of Eq.~\eqref{EQ:heatExch}~\cite{JPEl:2018, VPa:2013} 
\begin{equation}
\label{EQ:Sph_Cyl}
q_{1-2} = \frac{\sigma A_{2}(T_{1}^{4} - T_{2}^{4})}{\frac{1}{\varepsilon_{2}} + \frac{1-\varepsilon_{1}}{\varepsilon_{1}}\left(\frac{A_{2}}{A_{1}} \right)}~~.    
\end{equation}
The heat exchange between two concentric spheres, with $A_{1}$ and $A_{2}$ being the outer and 
inner surfaces respectively, and kept at temperatures $T_{1} > T_{2}$, such that the heat 
transfer takes place from out$-$to$-$inside, follows the same expression as in 
Eq.~\eqref{EQ:Sph_Cyl}~\cite{WEBlink}. The heat exchange for 
parallel flat plates of area $A = A_{1} = A_{2}$, $F_{21} = 1$, when $T_{1} > T_{2}$, is given by
\begin{equation}
\label{EQ:Parallel}
q_{1-2} = \frac{\sigma A(T_{1}^{4} - T_{2}^{4})}{\left(\frac{1}{\varepsilon_{1}} + \frac{1}{\varepsilon_{2}} -1 \right)}~,\mathrm{and~for~same~meterial}~(\varepsilon_{1} = \varepsilon_{2} = \varepsilon)~~ \equiv \frac{\sigma A(T_{1}^{4} - T_{2}^{4})}{\left(\frac{2}{\varepsilon} -1 \right)}.    
\end{equation}

It is obvious from Eqns.~\eqref{EQ:General},~\eqref{EQ:Sph_Cyl}, and~\eqref{EQ:Parallel}
that the heat transfer can be minimized by using the low emissivity materials. 
As emissivity depends on the material as well as surface finishing and 
cleanliness, clean and well$-$polished metallic surfaces have 
low emissivity, whereas non$-$metallic surfaces have higher emissivity. 
Thus, thermal shields (walls) in the cryostat are normally made up of properly 
polished metallic surfaces (Copper or Aluminium)~\cite{V_Parma}. Copper is 
the most promising material for this purpose because: (1) It has less emissivity 
as compare to Aluminium, after proper mechanical polishing~\cite{V_Parma}. 
(2) Rare to brittle even at very low temperatures~\cite{White}. (3) It 
can be stored in underground storage and quickly moved to the processing 
sites~\cite{balshaw}. (4) The amount of cryogenic liquid used to 
cool 1~kg of Copper is less in comparison with other cryostat 
materials~\cite{White}. (5) Irradiation of Copper will produce $^{60}$Co 
whose lifetime is less as compared to those which are produced by other 
materials~\cite{White}. \par

Furthermore, if one introduces a third intermediate plate between the two plates, 
whose temperature is the average of the temperatures $\left(T_{3} = \frac{T_{1} + T_{2}}{2}\right)$
of the two plates, with $\varepsilon_{31}$ and $\varepsilon_{32}$ being its 
corresponding temperature$-$dependent emissivities, the general form of the 
heat exchange through the walls can be expressed as
\begin{equation}
\label{EQ:Gene_Ins}
q_{1-2} = \frac{\sigma (T_{1}^{4} - T_{2}^{4})}{\frac{1-\varepsilon_{2}}{\varepsilon_{2}A_{2}} + \frac{1}{A_{2}F_{23}} + \frac{1-\varepsilon_{32}}{\varepsilon_{32}A_{3}} + \frac{1-\varepsilon_{1}}{\varepsilon_{1}A_{1}} + \frac{1}{A_{3}F_{31}} + \frac{1-\varepsilon_{31}}{\varepsilon_{31}A_{3}}}~~.    
\end{equation}
In case of parallel plates ($A_{1} = A_{2} = A_{3} = A$), $F_{23} = F_{31} = 1$ as 
schematized in figure~\ref{fig:third_layer}(a), this heat exchange would be reduced 
by a factor of two~\cite{V_Parma} by the insertion of the same material 
($\varepsilon_{31} = \varepsilon_{32} = \varepsilon_{3} = \varepsilon$)
\begin{equation}
\label{EQ:factTWO}
q_{1-2} = \frac{\sigma A(T_{1}^{4} - T_{2}^{4})}{2\left(\frac{2}{\varepsilon} -1 \right)}~~.    
\end{equation}
In the case of spherical as well as cylindrical geometries with $A_{1}>A_{3}>A_{2}$, whose 
schematic diagrams are shown in figure~\ref{fig:third_layer}(b) and (c), the general 
form of heat exchange (Eq.~\eqref{EQ:General}) would take place the following form
\begin{equation}
\label{EQ:SPHERE:CYL}
q_{1-2} = \frac{\sigma A_{2}(T_{1}^{4} - T_{2}^{4})}{\frac{1}{\varepsilon_{2}} + \left(\frac{1}{\varepsilon_{1}}-1\right)\frac{A_{2}}{A_{1}} + 2\left(\frac{1}{\varepsilon_{3}}-1\right)\frac{A_{2}}{A_{3}} + \frac{A_{2}}{A_{3}}}~~.    
\end{equation}
Thus, in the case of spherical and cylindrical geometries, the heat load is also 
reduced nearly by a factor of two due to the insertion of a third intermediate layer 
as in the case of a parallel plate geometry. \par

This is a valuable hint about how the radiation heat load in a cryostat can be 
decreased by inserting a new third layer (or more layers) in between the hot and cold 
walls of the cryostat. This leads the people to proceed towards the MLI technique. In 
the MLI technique, the radiation heat from the outer space strikes on its first 
reflective layer. A part of heat radiation is reflected from this reflective layer 
and remaining radiation energy heats the first layer of the spacer. With the increment 
in the temperature of this spacer layer, solid conduction, gas conduction and radiation 
start taking place via the spacer material to the next reflective layer. It follows 
that the temperature of the second reflective layer will further increase. The second 
reflective layer reflects a part of the incident radiation to the first spacer layer 
and transfers the remaining energy to the second spacer layer~\cite{Sutheesh:2018}. 
This process continues up to the last bottom layer and results in an immense reduction 
in heat load. The basic functioning of the MLI is represented in figure~\ref{fig:MLI_Function}. 

\section{Optimization of the layer density in MLI technique}
\label{sec:MLI-Tech}
Selecting a suitable number of MLI layers is quite crucial and affects 
its efficiency at a significant level. It depends on the thickness as well as 
material of reflective layers employed, the thickness of coated Aluminium, type of 
spacers, thickness of blanket, residual gas pressure, etc. Therefore, it doesn't 
allow for introducing as many layers as can fit in the available space. 
This is because of using more radiation shields within a fixed available space 
might introduce other possible ways for the heat exchange such as: (1) Radiation 
between the shields, because the thickness of spacers is reduced, (2) Heat exchange 
due to solid conduction between radiation shields through the spacers is 
introduced, and (3) Gas conduction due to the presence of gas molecules between 
the radiation shields and spacers. Therefore in every MLI system, an optimum 
layer density (number of layers/cm) is defined where the heat transfer is minimum. 
A schematic diagram of the MLI system along with the heat exchange via heat radiation, 
solid conduction, and gas conduction between two consecutive reflective layers 
is shown in figure~\ref{fig:MLI_Function}. \par 

Large number of radiation shields within a fixed available space are responsible for 
the increment in the thermal contact. Consequently, solid conduction increases within 
these radiation shields. Furthermore, a large number of radiation shield means increase 
of material and assembly costs in the experiment. It follows that the layer density 
must be optimized~\cite{Johnson:2010, Hedayat:2000}. This optimization can be performed 
by varying the distance between layers and adjusting the number of layers 
within a fixed thickness of the insulation blanket. After the optimization of layer 
density, one needs to optimize the thickness of the blanket having the constant 
layer density. This is because the layer density is the same for all thicknesses 
if all the conditions and materials remain the same. \par 

There are two important analytical approaches available to optimize the 
layer density: (1) Modification of Lockheed equation, and (2) Theoretical calculation 
developed for Layer$-$by$-$Layer approach by McIntosh. In the Modified Lockheed 
equation~\cite{lockheed:1974}, the original Lockheed equation is modified by 
accounting the conductivity term for spacers in the solid conduction term by McIntosh. 
This approach accounted for all three modes of heat exchange~\cite{Hedayat:2000}: 
\begin{equation}
\label{eq:total_Heat}
q_{\rm{total}} = q (\mathrm{radiation}) + q (\mathrm{solid~conduction}) + q (\mathrm{gas~conduction})~~.
\end{equation}
The generalized form of the Modified Lockheed equation provides an empirical form 
of the heat flux such as~\cite{JEF:2018}:
\begin{equation}
\label{Q_Heat}
q_{\rm{total}} = \frac{C_{R}~\varepsilon~(T_{h}^{4.67}-T_{c}^{4.67})}{N} + \frac{C_{S}~\bar{N}^{2.63}(T_{h}-T_{c}) (T_{h}+T_{c})}{2~(N+1)} + 
\frac{C_{G}~P~(T_{h}^{0.52}-T_{c}^{0.52})}{N}~~,
\end{equation}
where $\bar{N}$ is the layer density and $N$ represents the number of layers. The term 
$C_{S}$ is the solid conduction coefficient and it is a function of spacer's material. 
Symbol $C_{R}$ is the radiation coefficient and it is a function of reflector's material. 
The $C_{G}$ is the gas conduction coefficient and it is a function of radiation gas 
pressure between the layers. Here $T_{h}$ is the outer (hot) layer temperature (in K), 
$T_{c}$ is the inner (cold) layer temperature (in K), and $P$ symbolizes the residual 
gas pressure. The Lockheed Equations are essentially based on the data from MLI systems 
(mainly composed of double$-$Aluminized Mylar radiation shields with Silk$-$net spacers), 
which are tested using a flat plate but not the case of cylindrical calorimeter~\cite{CUNN:1971}. 
The Modified Lockheed equation can not be used for Layer$-$by$-$Layer approach because of 
the presence of $(N+1)^{\rm{th}}$ term in the denominator of Eq.~\eqref{Q_Heat}. \par

One another important approach is the physics$-$based expression developed by McIntosh for 
the theoretical calculation of an MLI system performance~\cite{MCint:1993}. This approach 
also accounts for all the above mentioned (Eq.~\eqref{eq:total_Heat}) three modes of the 
heat exchange like the Lockheed Equations. All of these modes of heat exchange vary 
from layer to layer but, the total heat flux remains constant across the whole MLI$-$blanket.
It follows that this approach applies to a complete MLI system rather than a Layer$-$by$-$Layer 
approach. In McIntosh approach, the heat radiation exchange term comprises of~\cite{Hedayat:2000}:
\begin{equation}
\label{Q_Radiation}
q_{\rm{radiation}} = \frac{\sigma (T_{h}^{4}-T_{c}^{4})}{\left(\frac{1}{\varepsilon_{h}} + \frac{1}{\varepsilon_{c}} -1\right)}~~,
\end{equation}
where $\varepsilon_{h}$ and $\varepsilon_{c}$ are the emissivities of the hot and 
cold surfaces, respectively. The heat exchange through the gas conduction term consists 
of~\cite{RJ:1959}:
\begin{equation}
\label{Q_Gas_Conv}
q_{\rm{gas~conduction}} = C_{G}P\alpha(T_{h}-T_{c})~~,
\end{equation}
where $C_{G}=K_{G}/P\alpha$, in which $K_{G}$ is the gas conduction (in Wm$^{-2}$K$^{-1}$), $P$ 
is the gas pressure (in pa), $C_{G} = [(\gamma+1)/(\gamma-1)][R/8\pi MT]^{\frac{1}{2}}$ and 
its value is 1.1666 for the air and 2.0998 for the Helium, $\alpha$ is the accommodation 
coefficient, $\gamma = C_{p}/C_{v}$, $R$ is the gas constant (8.314~kJ$-$mol$^{-1}$K$^{-1}$), $M$ 
is the molecular weight of gas (in kg$-$mol$^{-1}$) and $T$ is the temperature of vacuum gauge 
(usually $\sim$300~K)~\cite{Sutheesh:2018}. Lastly, the heat exchange via the solid conduction 
term represents:
\begin{equation}
\label{Q_Solid_Cond}
q_{\rm{solid~conduction}} = K_{S}(T_{h}-T_{c})~~,
\end{equation}
where $K_{S}=C_{S}fk/\Delta x$, in which $C_{S}$ is an empirical constant, $f$ is the relative 
density of the separator compared to solid material, $k$ is the separator material 
conductivity (in Wm$^{-1}$K$^{-1}$) and $\Delta x$ ( $\equiv (d_{o}-d_{i})$ difference between the 
outer ($d_{o}$) and inner ($d_{i}$) diameter of the specimen) is the actual thickness separator 
between reflectors (in meter)~\cite{hedayat:2002}. Therefore the total heat transfer in McIntosh's 
approach comes out to be the sum of Eqns.~\eqref{Q_Radiation},~\eqref{Q_Gas_Conv}~and~\eqref{Q_Solid_Cond}:  
\begin{equation}
\label{McIntosh}
q_{\rm{total}} = \frac{\sigma (T_{h}^{4}-T_{c}^{4})}{\left(\frac{1}{\varepsilon_{h}} + \frac{1}{\varepsilon_{c}} -1\right)} + C_{S}fk\left(\frac{T_{h}-T_{c}}{\Delta x}\right) + C_{G}P\alpha(T_{h}-T_{c})~~.
\end{equation}

It is obvious from the Modified Lockheed equation~\eqref{Q_Heat} and McIntosh 
equation~\eqref{McIntosh} that the optimization of the layer density ($\bar{N}$) requires 
the selection and declaration of coefficients $C_{S}$, $C_{G}$ and $C_{R}$, whose values 
depend on the type of materials used. Reflector material is responsible for the radiation 
between shields and radiant heat transfer is proportional to the $\varepsilon$ of shields. 
It follows that low $\varepsilon$ materials must be used to form reflectors~\cite{Miller:2016}. 
There are two most frequent applicable reflectors: (1) Thin Aluminium$-$foil, and (2) 
Polymeric film composed of polyester (Mylar) coated with Aluminium on both sides called 
Double$-$Aluminized$-$Mylar; or composed by Polyimide (Kapton)~\cite{Miller:2016, EURECA:2014}. \par 

The spacer material is responsible for the solid conduction between reflector sheets through 
spacers. Furthermore, the heat transfer through spacers is proportional to the thermal 
conductivity of the material. Therefore, low conductivity materials like Silk$-$net, 
Polyester$-$net, Fiberglass cloth, Fiberglass mats, Paper, Rayon fiber paper, etc. are 
used to form spacers~\cite{JEF:2018}. \par

In case of imperfect vacuum insulation, the residual gaseous conduction can develop 
a nontrivial contribution to the total heat transfer. It depends on the level of vacuum, the 
type and amount of residual gas, geometry, and temperatures involved~\cite{JRF:2012}. 
Generally, the level of vacuum between layers is unknown and can have considerable effects 
on the thermal performance of the MLI system. It should be noted that the residual gas 
conduction is proportional to the pressure and temperature difference ($T_{h}-T_{c}$) 
between walls~\cite{V_Parma}. Thus it can be minimized by lowering the pressure of the gas 
(P $\lesssim10^{-4}$ torr). To account for and explain all the thermal performance 
results of MLI system, it requires an understanding of all the available information 
from the heating, purging, evacuation, and vacuum monitoring steps~\cite{JEF:2018}. \par

Although one can use more radiation shields to reduce the radiation, the solid conduction 
between radiation shields increases through the spacers due to the decreased space between 
two radiation shields and the thickness of spacers. It follows that there must be a balance 
point between the radiation and solid conduction heat load, which is also called as optimum 
layer density. In the following section, we will elaborate the optimization of layer density 
for achieving the minimum heat transfer. \par

\section{Results and discussion}
\label{sec:results_discus}

A stable temperature of the cryogenic liquid can be achieved by reducing the heat load coming 
from the outer wall of the cryostat. Numerous approaches have been tested and are being utilized 
to reduce the heat load. An optimum level of vacuum is required between the two walls of the 
cryostat to reduce the gaseous conduction heat load. Furthermore, in order to reduce the loss 
of cryogenic liquid due to evaporation, the silver coating has also been done at the inner 
surface of the outer wall and the outer surface of the inner wall, but still, heat load persists 
at significant level~\cite{EURECA:2014}. The present work will not go into the detail of the 
optimization of the vacuum level between the walls of the cryostat and the choice of suitable 
metallic coatings over them. This work is focused on the optimization of the MLI technique for 
the reduction of thermal radiation heat load to the cryogenic liquids.

\subsection{MLI testing material}
\label{subsec:material_disc}

The current work assumes that the radiation shields and spacers in the MLI technique are 
placed perpendicular to the direction of the heat flow. These spacers are placed to avoid 
the thermal contact between the radiation shields because this causes the production of 
conductive heat load. In the testing of cryostat's thermal performance, the $\varepsilon$ 
of the outer surface is chosen to be 0.043 for Aluminized radiation shields, optimum vacuum 
pressure $P$~=~10$^{-4}$ torr, residual gas is Nitrogen, and the cold and hot boundary 
temperatures are approximately 77 K for LN$_{2}$ and 300 K for water, respectively. This 
level of vacuum is required to minimize the conductive heat load produced by the presence 
of residual Nitrogen gas in between the walls of the cryostat. Furthermore, this partial 
conduction may also cause the condensation of moisture around the outside of the 
walls, which may further produce undesirable heat load on the cryogenic 
liquid~\cite{janis:2009}. \par

The present analysis follows the robust Modified Lockheed equation~\eqref{Q_Heat} to 
evaluate the production of heat load in the MLI technique. The coefficients $C_{R}$, 
$C_{G}$, and $C_{S}$ used in this expression are material dependent. We have selected: 
(1) Unperforated double Aluminized Mylar sheet (DAM) with Silk$-$net, (2) Perforated 
DAM with Glass$-$tissue, and (3) Perforated DAM with Dacron materials as reflecting 
shields as well as spacers in the MLI technique. The values of coefficients $C_{R}$, 
$C_{G}$, and $C_{S}$ for these selected materials are listed in table~\ref{ref:material}. \par

\begin{table}[h!]
\begin{center}
\setlength{\tabcolsep}{0.16em} 
\renewcommand{\arraystretch}{1.1}
\centering
\caption{Selected reflective layer as well as spacer materials for the MLI technique with their 
  respective coefficients.}
\vspace*{8pt}
\label{ref:material}
{\def\arraystretch{1}\tabcolsep=3pt}
\begin{tabular}{|c|c|c|c|} \hline
                                                        & \multicolumn{3}{|c|}{{Coefficients}}                                     \\ \cline{2-4} 
~Materials~                                             &~$C_{R}$~            &~$C_{S}$~                      &~$C_{G}$~             \\ \cline{1-4} 
~Unperforated DAM with Silk$-$net~~~~~~\cite{JEF:2018}~&~5.39$\times10^{-10}$~&~8.95$\times10^{-8}$~          &~1.46$\times10^{-4}$~ \\ \hline
~Perforated DAM with Glass$-$tissue~~\cite{Hedayat:2000}~&~7.07$\times10^{-10}$~&~7.30$\times10^{-8}$~          &~1.46$\times10^{-4}$~ \\ \hline
~Perforated DAM with Dacron ~~~~~~~~~~\cite{Hedayat:2000}~&~4.94$\times10^{-10}$~&~Shown in~\eqref{Eq_constants:Mod}~  &~1.46$\times10^{-4}$~ \\ \hline  

\end{tabular}
\end{center}
\end{table}

The ``solid conduction'' term in the empirical expression~\eqref{Q_Heat} would be modified 
for the Dacron spacer material. It has been done because in the original Lockheed equation 
the spacer material used was Glass$-$tissue with different sizes of shield's perforation, 
whereas during testing the spacer used was Dacron material with dissimilar sizes of 
shield's perforation. It follows that the term containing ``solid conduction'' in~\eqref{Q_Heat} 
got modified for the Dacron material~\cite{Hedayat:2000}. After incorporating this modification, 
the total heat flux expression for the Dacron becomes

\begin{eqnarray}\nonumber
\label{Eq_constants:Mod}
q_{\rm{total}} = \left(\frac{2.4\times10^{-4}(0.017 + 7\times10^{-6}~(800-T) + 0.0228~ln(T))~\bar{N}^{2.63}(T_{h}-T_{c})}{N}\right) \\
+ \left(\frac{C_{R}~\varepsilon~(T^{4.67}_{h}-T^{4.67}_{c})}{N}\right) + \left(\frac{C_{G}~P~(T^{0.52}_{h}-T^{0.52}_{c})}{N}\right)~, 
\end{eqnarray}
where $T$ is the average temperature $\left(T = \frac{T_{h} + T_{c}}{2}\right)$ of the 
hot and cold boundaries. It is obvious from Eq.~\eqref{Eq_constants:Mod} that the term 
containing solid conduction segment is modified for the Dacron~\cite{Hedayat:2000}. We 
will proceed with these important and necessary ingredients towards the parameter analysis.

\subsection{Analysis of the key parameters in MLI technique}
\label{subsec:key_analysis}
Apart from the selection of suitable radiation shield as well as spacer materials, the 
Layer density, number of layers, and thickness of insulating blanket, are the key 
parameters in getting excellent MLI with potential performance in reducing the thermal
radiation heat load. In the present work, we have calculated the effect of layer density 
and the number of layers on the heat load and evaluated optimal layer density and favorable 
thickness of insulating blanket for these selected materials in the MLI technique. \par

\subsubsection{Enhancement in layer density and its effect on the heat load}
\label{subsubsec:test_heat_load}
\begin{figure}[h!]
  \begin{center}
    \includegraphics[width=11.0cm]{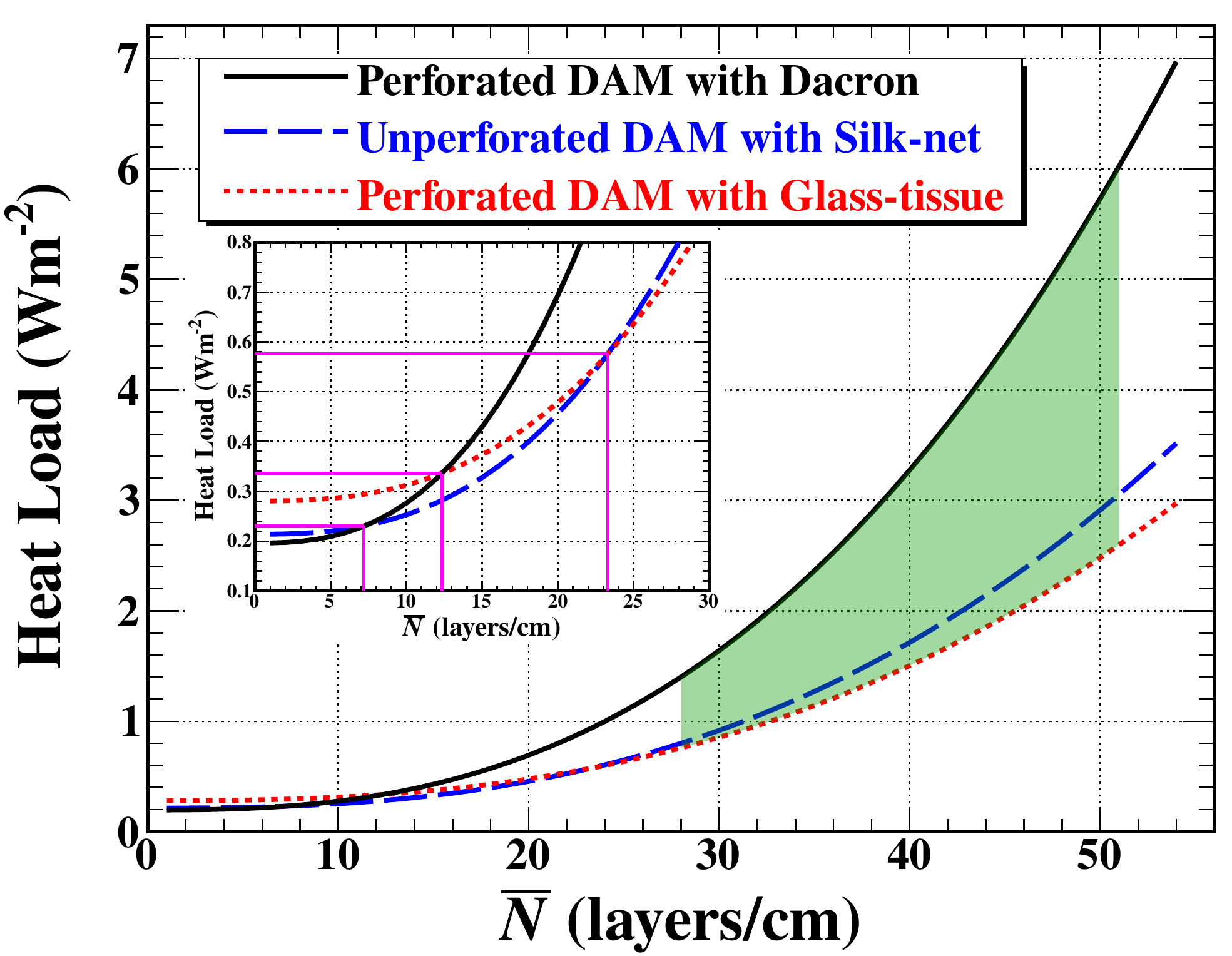}
  \end{center}
  \caption{Variation in the heat load with the increment in $\bar{N}$ for a constant 
    number of layers ($N = 40$). The effect of smaller $\bar{N}$ choice on the heat 
    load and the comparative material's behavior is shown in the inset.}
\label{fig:enh_LD}
\end{figure}

Radiation shields are generally coated with highly reflecting metals (Aluminium or Gold),
to reflect the thermal radiation for reducing the radiation heat load. It follows that 
there may be a chance of conduction between them and may lead to the enhancement in the 
heat load. Thus, insulating spacers are used in between the radiation shields to solve this 
issue. Although the installation of insulating spacers causes the decrement in heat load, 
it doesn't reduce to a significant level. Therefore, more and more radiation shields 
with insulating spacers can be used to reduce the heat load. However, with the increment in 
$\bar{N}$, the heat load starts increasing due to the reduction of space between the 
radiation shields and thus solid conduction between the shields increases through the 
spacers~\cite{V_Parma}. This increment in the heat load with $\bar{N}$ for a fixed 
number of layers ($N = 40$ is chosen as the reference for explanation), is shown in 
figure~\ref{fig:enh_LD}.

It is obvious from figure~\ref{fig:enh_LD} that the heat load increases with $\bar{N}$ 
for all the three selected materials. In the light of Eq.~\eqref{Eq_constants:Mod} and 
coefficients of table~\ref{ref:material}, this increment in the heat load is quite 
larger for the Dacron in comparison to the Glass$-$tissue and Silk$-$net material at higher 
$\bar{N}$. However, below $\bar{N} \sim10$ the Dacron material leads to the minimum heat 
load among the three selected materials. It is also noteworthy in the inset of 
figure~\ref{fig:enh_LD} that the perforated DAM with Glass$-$tissue and unperforated DAM 
with Silk$-$net for $\bar{N} = 23.3$ produces equivalent heat load of 0.58 wm$^{-2}$. 
Similarly, perforated DAM with Dacron and perforated DAM with Glass$-$tissue at 
$\bar{N} = 12.3$ leads to the concordant heat load of 0.34 wm$^{-2}$. Furthermore, 
perforated DAM with Dacron and unperforated DAM with Silk$-$net produces comparable heat 
load of 0.23wm$^{-2}$ at $\bar{N} = 7.2$. \par
\begin{table}[h!]
\begin{center}
\setlength{\tabcolsep}{0.16em}
\renewcommand{\arraystretch}{1.1}
\centering
\caption{Enhancement in the heat load with the increment in $\bar{N}$ (from 28 to 51 
  layers/cm $\equiv$ 82\%) for the chosen reflective layer and spacer materials.}
\vspace*{8pt}
\label{perc:increment}
{\def\arraystretch{1}\tabcolsep=3pt}
\begin{tabular}{|c|c|c|c|} \hline
~Reflective layer~   &~Spacer material~ &~Increment in $\bar{N}$ (\%)~  &~Enhancement in heat load (\%)  \\ \hline 
~Unperforated DAM~   &~Silk$-$net~      &                               &~282                            \\ \cline{1-2} \cline{4-4}     
~Perforated DAM~     &~Glass$-$tissue~  &~82~                           &~242                            \\ \cline{1-2} \cline{4-4}  
~Perforated DAM~     &~Dacron~          &                               &~330                            \\ \hline

\end{tabular}
\end{center}
\end{table}
The effect of enhancement in the layer density from $\bar{N} = 28$ to 51 layers/cm 
($\equiv$ 82\% increment, as a reference) on the heat load for the chosen materials is 
summarized in table~\ref{perc:increment}. It is obvious from the shaded region of 
figure~\ref{fig:enh_LD} that this increment in $\bar{N}$ leads to severe enhancement 
in the heat load. This enhancement is even more profound for the perforated DAM with 
Dacron material in comparison to the unperforated DAM with Silk$-$net and perforated 
DAM with Glass$-$tissue. This drastic enhancement in the heat load marks that the 
value of $\bar{N}$ can not be increased beyond a limit and must need to be optimized 
for different materials.  

\subsubsection{Selection of optimal layer density}
\label{subsubsec:optimal_LD}
The optimization of $\bar{N}$ has a great impact on achieving the better thermal 
performance of the MLI technique. The value of $\bar{N}$ varies for different 
materials~\cite{WEBlink}. Radiation shields and spacers are used to reduce the 
radiation heat load. Consequently, this heat load reduces however, to minimize it 
significantly one may increase the number of radiation shields. A huge increment 
in the number of radiation shields (and thus $\bar{N}$) lead the probable enhancement 
in the thermal contact between the radiation shields because of the decrement in 
spacer thickness as well as the space between radiation shields. This results in
an increment in the solid conduction between the radiation shields through the 
spacers. Whereas, according to Eq.~\eqref{Q_Heat}, the radiation heat load remains 
invariant for a constant value of $N$. It follows that there must be an equilibrium 
between the radiation and conduction heat loads. This equilibrium point would 
represent the optimal $\bar{N}$ where the total heat load is minimum for that 
material~\cite{V_Parma,Johnson:2010}. \par

\begin{figure}[h!]
  \begin{center}
    \includegraphics[width=11.0cm]{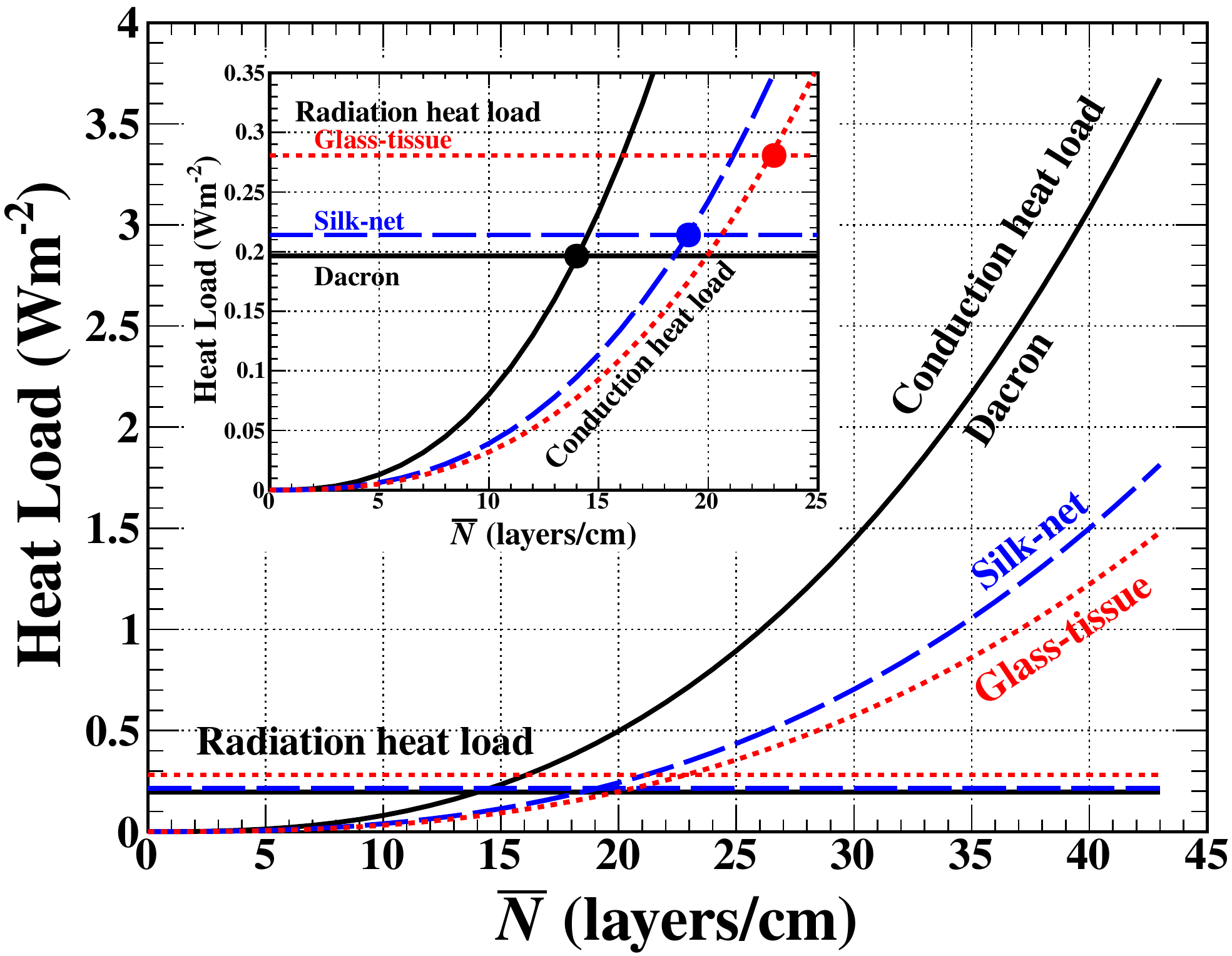}
  \end{center}
  \caption{Variation in the heat load with $\bar{N}$ for the chosen unperforated DAM 
    with Silk$-$net, perforated DAM with Glass$-$tissue and perforated DAM with Dacron 
    materials in the MLI technique. The intersecting contours (as shown in the inset) 
    of radiation and conduction heat load represents the optimum value of $\bar{N}$ 
    for reference value of $N$ = 40.}
\label{fig:Opt_LD}
\end{figure}

We have calculated the optimal $\bar{N}$ for the selected perforated DAM with 
Glass$-$tissue, perforated DAM with Dacron, as well as unperforated DAM with the Silk$-$net 
radiation shield and spacer materials for a reference $N$ = 40. At this reference 
value of $N$, the variation of radiation and conduction heat load with $\bar{N}$ is 
displayed in figure~\ref{fig:Opt_LD}. Although the value of radiation heat load 
varies with the choice of material, remain constant for a selected constant value 
of $N$ and shown by the horizontal contours for the selected materials. The 
equilibrium between the radiation and conduction heat load is zoomed and shown in 
the inset of figure~\ref{fig:Opt_LD}. The intersection points (represented by solid 
dots) in the variation of conduction and radiation heat load with $\bar{N}$ are the 
optimum values of $\bar{N}$ for the respective materials. The optimum value 
of $\bar{N}$ for a reference value $N$ = 40 is summarized in table~\ref{tab:Opt:LD} 
for the selected materials. If all the above mentioned criterion remains intact, the 
MLI insulating blanket's thicknesses would be the same for a given $\bar{N}$. It 
follows that the thickness of the insulating blanket needs to be optimized. As a 
consequence of the optimization in $\bar{N}$, the optimal thickness of the insulating 
blankets are also evaluated. These values are listed in the last column of 
table~\ref{tab:Opt:LD}. \par

\begin{table}[h!]
\begin{center}
\setlength{\tabcolsep}{0.16em}
\renewcommand{\arraystretch}{1.1}
\centering
\caption{Optimal values of $\bar{N}$ and the thickness of insulating blanket for the 
  selected unperforated DAM with Silk$-$net, perforated DAM with Glass$-$tissue and 
  perforated DAM with Dacron materials for a reference value $N$ = 40.}
\vspace*{8pt}
\label{tab:Opt:LD}
{\def\arraystretch{1}\tabcolsep=3pt}

\begin{tabular}{|c|c|c|c|c|} \hline
~Reflective layer~   &~Spacer material~ &~Reference $N$~  &~$\bar{N}$ (layers/cm) &~Blanket's thickness (mm)~   \\ \hline                 
~Unperforated DAM~   &~Silk$-$net~      &                 &~19.0                  &~21.1~                       \\ \cline{1-2} \cline{4-5}    
~Perforated DAM~     &~Glass$-$tissue~  &~40~             &~22.7                  &~17.6~                       \\ \cline{1-2} \cline{4-5} 
~Perforated DAM~     &~Dacron~          &                 &~14.0                  &~28.6~                       \\ \hline

\end{tabular}
\end{center}
\end{table}

The outcome of table~\ref{tab:Opt:LD} exhibits that the perforated DAM with Glass$-$tissue 
can accommodate larger $N$ within a fixed space between the cold and hot boundaries of 
cryostat in comparison to the unperforated DAM with Silk$-$net and perforated with DAM Dacron. 
Therefore, perforated DAM with Glass$-$tissue would be a better choice with optimal $\bar{N}$ 
of 22.7 layers/cm, and optimal thickness 17.6~mm of blanket for a constant $N$ = 40 in 
making radiation shield and spacer for MLI technique. However, instead of a constant $N$, 
the effect of variable $N$ on heat load needs to be considered before coming to the 
inference in the selection of material for MLI technique.

\begin{figure}[h!]
  \begin{center}
    \includegraphics[width=11.0cm]{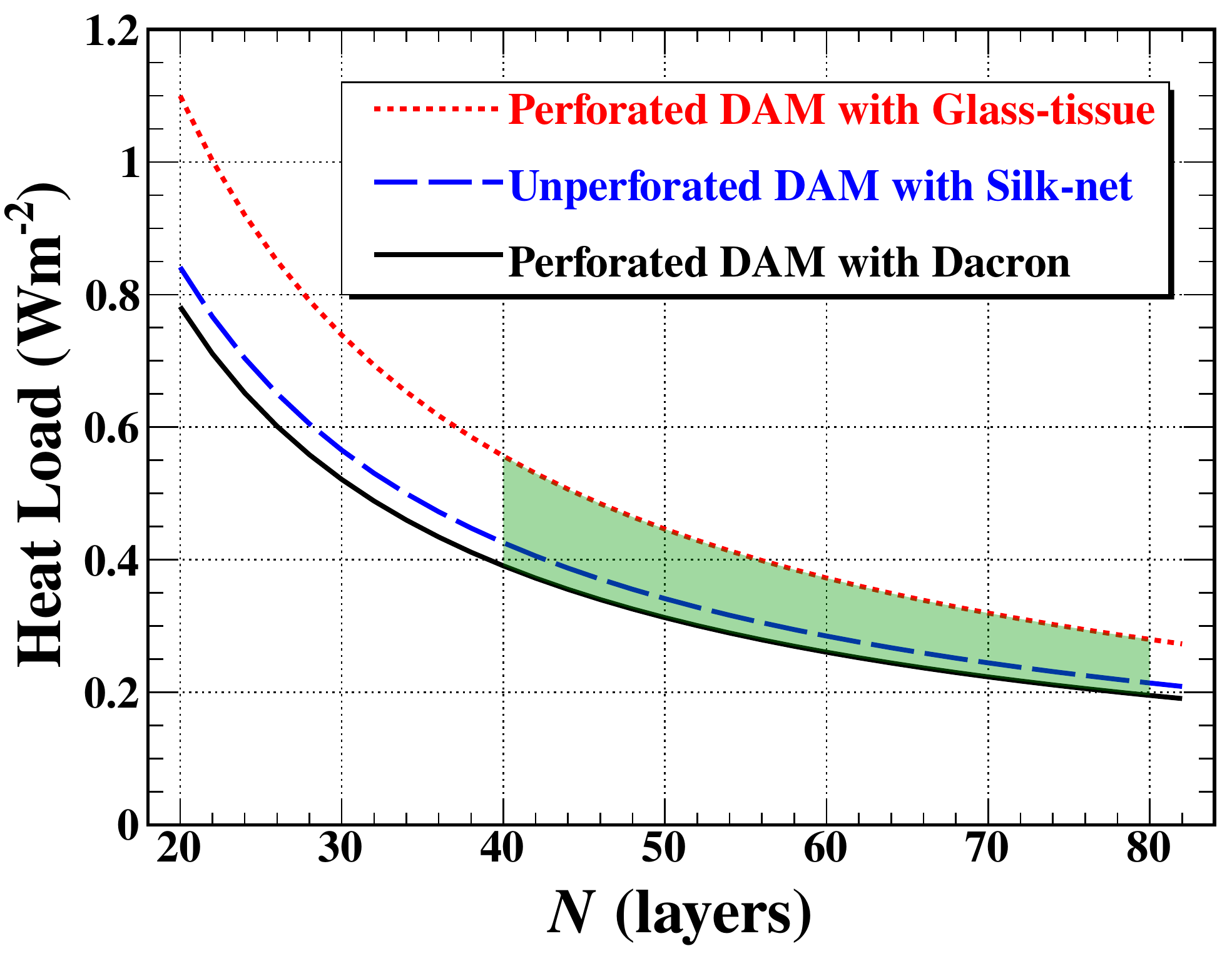}
  \end{center}
  \caption{Effect of increment in the values of $N$ over the heat load at the calculated 
    optimum values of $\bar{N}$ for the chosen materials. As a reference, the effect of 
    an increment in $N$ from 40 to 80 is shown by the shaded region.}
\label{fig:IncH:HLD}
\end{figure}

\subsubsection{Effects of increment in the number of layers over the heat load}
\label{IncN:HLD}
After the evaluation of optimum values of $\bar{N}$ (via variation between heat load and 
$\bar{N}$), it becomes realistic to further quantify the effect of increment in $N$ over 
the heat load. One has to increase the thickness of the MLI system for making increments in 
the value of $N$ at the optimum and constant value of $\bar{N}$. This would result in a 
further decrement in the heat load because of the reduction in the thermal contact 
between the radiation shields. \par

\begin{table}[ht]
\begin{center}
\setlength{\tabcolsep}{0.16em}
\renewcommand{\arraystretch}{1.1}
\centering
\caption{The expected decrement in the heat load with enhancement in the value of $N$ at the 
  constant and optimum values of $\bar{N}$ for the chosen materials in MLI technique.}
\vspace*{8pt}
\label{Decr:Heat_LD}
{\def\arraystretch{1}\tabcolsep=3pt}
\begin{tabular}{|c|c|c|c|} \hline
~Reflective layer~   &~Spacer material~  &~Increment in $N$ &~Decrement in heat load (\%)~  \\ \hline  
~Unperforated DAM~   &~Silk$-$net~       &                  &~50.0~                         \\ \cline{1-2} \cline{4-4}   
~Perforated DAM~     &~Glass$-$tissue~   &~40~$-$~80~       &~50.0~                         \\ \cline{1-2} \cline{4-4}  
~Perforated DAM~     &~Dacron~           &                  &~50.0~                         \\ \hline

\end{tabular}
\end{center}
\end{table}

Knowing the optimal values of $\bar{N}$, we have further investigated the effect 
of increment in the $N$ at the constant values of $\bar{N}$ for the selected materials.
A variation of the heat load with $N$ at constant $\bar{N}$ is shown in 
figure~\ref{fig:IncH:HLD}, which exhibits an expected decrement in the heat load with $N$. 
In order to get the inference about the performance of the material, we have evaluated 
the decrement in heat load with an increment of 100\% in $N$ from 40 to 80 as shown by 
the shaded region in figure~\ref{fig:IncH:HLD} and their outcomes are summarized in 
table~\ref{Decr:Heat_LD} for the chosen materials.  

It is obvious from table~\ref{Decr:Heat_LD} and figure~\ref{fig:IncH:HLD} that an increment 
in $N$ by 100\% (at the constant values of $\bar{N}$ taken from table~\ref{tab:Opt:LD}) 
causes for 50\% decrement in heat load. As this decrement in heat load is evaluated at the 
respective optimum values of $\bar{N}$, for the selected materials, the decrement 
in heat load is exactly the same and material independent. 

\subsection{Preference in the geometry of the cryostat}
\label{Comp:Cryostat}
Cryostats are being used with different perspectives in different experiments. They need
to accommodate the cryogenic liquid along with the shielding around the detector, detectors 
immersed in the cryogenic liquid, and only cryogenic liquid, which also works as the detector, 
etc. Cylinders, parallel flat plates, and spheres with two enclosed envelopes are of particular 
interest in the cryostat design. Apart from taking care of several aspects like the choice of 
material, selection of $N$, optimization of $\bar{N}$, the thickness of insulating blankets, 
appropriate geometry, the problem of buckling, etc., there is one common requirement in such 
cryostats is that, they all are intended to minimize their heat load. \par
\begin{figure}[h!]
  \begin{center}
    \includegraphics[width=11.0cm]{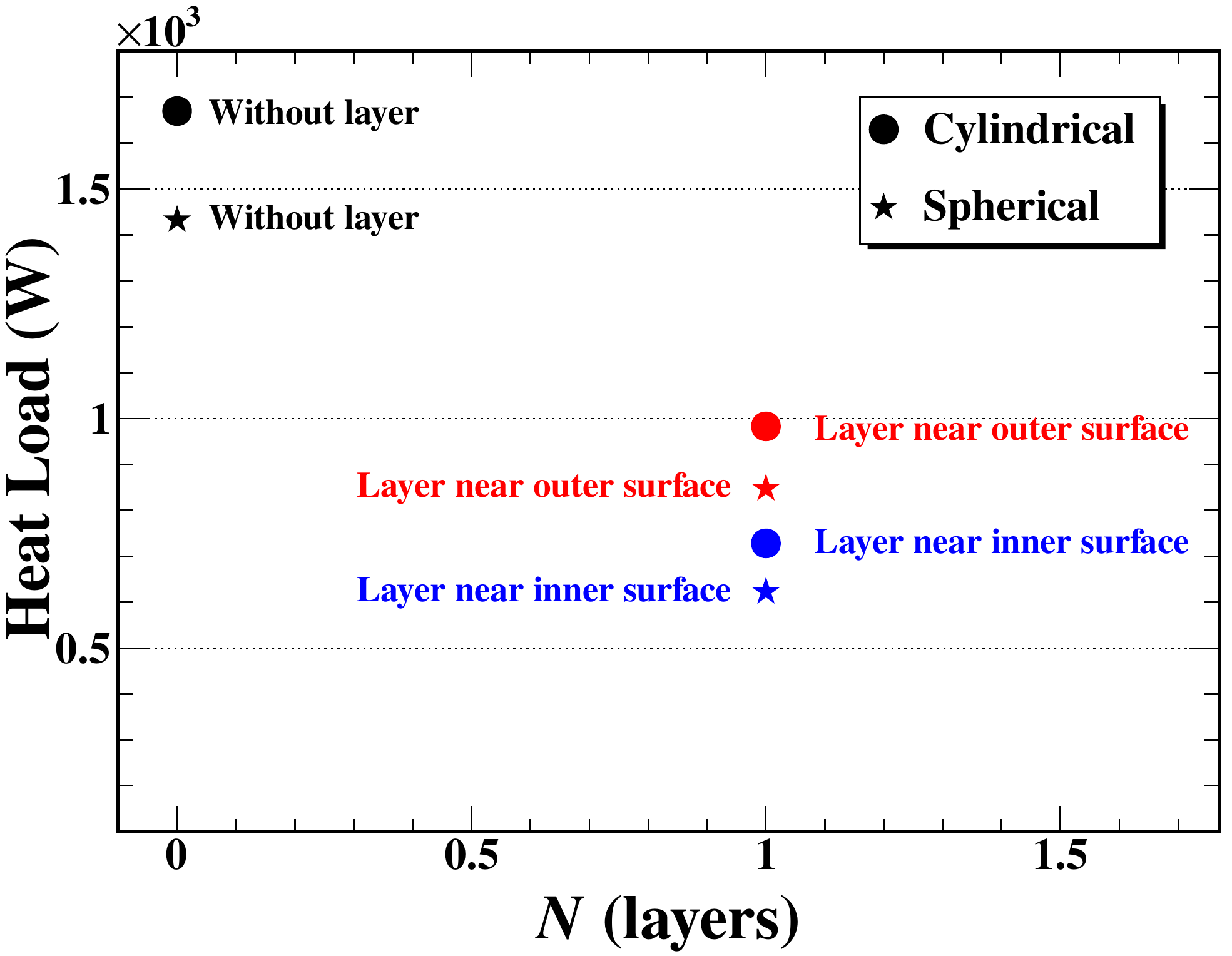}
  \end{center}
  \caption{Comparison of the heat load production before and after insertion of one MLI layer 
    near inner as well as outer surfaces in the spherical and cylindrical geometries of the 
    cryostat.}
\label{fig:comp_SP_CY}
\end{figure}
The radiation exchange between two surfaces depends on their geometry, which can be well 
understood by the geometrical view factor~\cite{MFMd:2013, WEB}. Therefore the geometry of a 
cryostat would need to be better designed under proper consideration of the total heat load. 
It follows that the analysis of all the key parameters discussed above in 
section~\ref{subsec:key_analysis} would have a significant impact on the selection of 
appropriate geometry for the cryostat in an experiment. \par

There are spherical, cylindrical and conical geometries of a cryostat, which are very common 
in experiments. The present work is focused on the comparison of cylindrical and spherical 
cryostats (with exactly the same volume) in the minimization of heat load. We have considered 
the dimensions of GERDA cryostat for comparison of geometries~\cite{GERDA:2005, GERDA:2006}. 
A variation of heat load with a number of MLI layers for both of these geometries is shown in 
figure~\ref{fig:comp_SP_CY}. It is explicit from figure~\ref{fig:comp_SP_CY} that the 
spherical cryostats are comparatively more suitable than the cylindrical ones in holding the 
cryogenic liquid for a longer time due to the less heat load. Therefore, spherical geometry is 
the most effective configuration because of having less heat load due to the least surface 
area to volume ratio. However, cylindrical cryostats are more preferably being used in 
several experiments because the fabrication of the cylindrical geometry is relatively easy 
and it is the most economical configuration in comparison to the spherical or conical 
geometries~\cite{thapa:2013}. \par

The effect of the MLI approach has a significant impact on reducing the heat load, which is 
obvious from figure~\ref{fig:IncH:HLD}. In the MLI technique, layering near the inner surface 
reduces heat load nearly by a factor of two than layering close to the outer surface. This 
reduction in heat load by MLI technique is quite similar for both spherical as well as 
cylindrical geometries. \par

\section{Summary and conclusion}
\label{sec:summary}
MLI is an essential and important insulating technique used worldwide in the field of cryogenic 
as well as in the space industry. The efficient as well as long time storage, transfer of 
cryogens, cooling of scientific instruments, and exploration of space demand a robust MLI 
technique in thermal insulation systems. It follows that a tremendous investigation is being 
performed since the last two decades in the field of MLI for a range of environments from high 
vacuum to no vacuum. A vigorous MLI system following a ``thermo$-$economic'' approach must need 
to consider the design of cryostat, appropriate radiation shield as well as spacer material, 
transmissivity of MLI, interstitial gas and its pressure, structural supports, and other 
mechanical obstacles to achieve the best MLI system. \par

The present work has analyzed the performance of perforated DAM with Dacron, unperforated DAM with 
Silk$-$net and perforated DAM with Glass$-$tissue as the radiation shield as well as spacer materials in 
MLI technique. This analysis follows the robust Modified Lockheed equation~\eqref{Q_Heat} to calculate 
the production of heat load in the MLI technique with these materials. In the present work, we 
have calculated the effect of layer density and number of layers on the heat load and 
evaluated optimal layer density and favorable thickness of insulating blanket for these 
selected materials in MLI technique. \par

It is observed that the increment in layer density causes the increment in heat load due 
to the reduction of space between the radiation shields and thus solid conduction between the 
shields increases through the spacers. This enhancement in heat load is significantly larger 
for the perforated DAM with Dacron and comparatively smaller for perforated DAM with Glass$-$tissue. With
an increment in layer density from 28 to 51 layers/cm for a constant 40 insulating layers, the 
heat load increases by 330\% in case of perforated DAM with Dacron, whereas this increment in the 
heat load is around 242\% for perforated DAM with Glass$-$tissue. This huge enhancement in the heat 
load reveals the necessity for the optimization of layer density. \par

According to the Modified Lockheed equation~\eqref{Q_Heat}, the radiation heat load remains 
constant whereas, the conduction heat load increases with layer density for a selected constant 
number of insulating layers. Therefore, there must be an equilibrium point between the radiation 
and conduction heat loads. The total heat load is minimum at this equilibrium point for that 
material and thus provides the optimal layer density. The optimal layer density for perforated DAM 
with Glass$-$tissue comes out to be 22.7 layers/cm (optimal blanket's thickness = 17.6 mm) which is 
the best among the three selected materials and the most conservative 14.0 layers/cm (optimal 
blanket's thickness = 28.6 mm) for the perforated DAM with Dacron. Furthermore, an increment of 
100\% in the number of layers at these optimal layer densities for the respective materials leads 
to 50\% decrement in the heat load. Therefore, perforated DAM with Glass$-$tissue would be a better 
choice by optimal layer density and thickness of blanket in making radiation shield and spacer for 
getting the lowest heat load among the three chosen materials in the MLI system. \par

The inference of this analysis has a significant impact on the selection of appropriate 
geometry for the cryostat in an experiment. In the present, work we have compared the 
effect of MLI technique on the heat load in cylindrical and spherical cryostats. It comes 
out that, in comparison to the cylindrical cryostat, the spherical cryostat leads to the 
less heat load and thus more suitable in holding the cryogenic liquid for a longer time. 
The MLI technique has a significant impact on reducing the heat load and a single layering near 
the inner surface of cryostat reduces heat load nearly by factor of two than the layering 
close to the outer surface.  \par

\acknowledgments
The author D. Singh gratefully acknowledges Council of Scientific and Industrial Research 
(CSIR-UGC), New Delhi, India, for the financial support in the form of CSIR (JRF/SRF) 
fellowship. The authors D. Singh, A. Pandey, and V. Singh are thankful to the Ministry of 
Human Resource Development (MHRD), New Delhi, India for the financial support through 
Scheme for Promotion of Academic and Research Collaboration (SPARC) project 
No. SPARC/2018$-$2019/P242/SL.

\end{document}